# On-Demand Multicasting in Ad-hoc Networks: Performance Evaluation of AODV, ODMRP and FSR

Rajendiran. M[1][*]    Srivatsa. S. K[2]

[1] Department of Computer Science and Engineering, Sathyabama University, Chennai, Tamilnadu 600119, India

[2] Department of Computer Science and Engineering, St. Josephs College of Engineering, Chennai, Tamilnadu 600119, India

**Abstract**
Adhoc networks are characterized by connectivity through a collection of wireless nodes and fast changing network topology. Wireless nodes are free to move independent of each other which makes routing much difficult. This calls for the need of an efficient dynamic routing protocol. Mesh-based multicast routing technique establishes communications between mobile nodes of wireless adhoc networks in a faster and efficient way.
In this article the performance of prominent on-demand routing protocols for mobile adhoc networks such as ODMRP (On Demand Multicast Routing Protocol), AODV (Adhoc on Demand Distance Vector) and FSR (Fisheye State Routing protocol) was studied. The parameters viz., average throughput, packet delivery ration and end-to-end delay were evaluated. From the simulation results and analysis, a suitable routing protocol can be chosen for a specified network. The results show that the ODMRP protocol performance is remarkably superior as compared with AODV and FSR routing protocols.

## 1. Introduction

One of the basic internet tasks is routing between various nodes. It is nothing other than establishing a path between the source and the destination. However in large and complex networks routing is a difficult process because of the possible intermediate hosts it has to cross in reaching its final destination. In order to reduce the complexity, the network is considered as a collection of sub domains and each domain is considered as a separate entity. This helps routing easy [1]. However basically there are three routing protocols in ad hoc networks namely proactive, reactive and hybrid routing protocols. Of these reactive routing protocols establish and maintain routes based on demand.

The reactive routing protocols (e.g. AODV) usually use distance-vector routing algorithms that keep only information about next hops to adjacent neighbors and costs for paths to all known destinations [2]. The reactive routing protocols (e.g. AODV) usually use distance-vector routing algorithms that keep only information about next hops to adjacent neighbors and costs for paths to all known destinations [2].

On the other hand hybrid routing protocols combine the advantages of both proactive and reactive protocols. Reliable multicast in mobile network was proposed by Prakash et al, [3]. In their solution the multicast message is flooded to all the nodes over reliable channels. The nodes then collectively ensured that all mobile nodes belonging to the multicast group get the message. If a node moves from one cell to another while a multicast is in progress, delivery of the message to the node was guaranteed.

Tree-based multicast routing provides fast and most efficient way of routing establishment for the communications of mobile nodes in MANET [4]. The authors described a way to improve the throughput of the system and reduce the control overhead. When network load increased, MAODV ensures network performance and improves protocol robustness. Its PDR was found to be effective with reduced latency and network control overhead. On Demand Multicast Routing Protocol is a multicast routing protocol(ODMRP) designed for ad hoc networks with mobile hosts [5]. Multicast is nothing but communication between a single sender and multiple receivers on a network and it transmits a single message to a select group of recipients [6]. Multicast is commonly used in streaming video, in which many megabytes of data are sent over the network. The major advantage of multicast is that it saves bandwidth and resources [7]. Moreover multicast data can still be delivered to the destination on alternative paths even when the route breaks. It is an extension to Internet architecture supporting multiple clients at network layers. The fundamental motivation behind IP multicasting is to save network and bandwidth resource via transmitting a single copy of data to reach multiple receivers. Single packets are copied by the network and sent to a specific subset of network addresses. These addresses point to the destination. Protocols allowing point to multipoint efficient distribution of packets are frequently used in access grid applications. It greatly reduces the transmission cost when sending the same packet to multiple destinations.

A primary issue in managing multicast group dynamics is the routing path built for data forwarding. Most existing ad hoc multicasting protocols can be classified as tree-based or mesh-based. The tree-based protocol, a tree-like data forwarding path is built with the root at the source of the multicast session. The mesh-based protocol [eg. ODMRP], in contrast, provide multiple routes between any pair of source and destination, intended to enrich the connectivity among group members for better resilience against topology changes.

## 2. Literature Survey

A lot of work has been done to evaluate the performance of routing protocols in ad hoc networks. Thomas Kunz et al. [8] compared AODV and ODMRP in Ad-Hoc Networks. Yadav et al. [9] studied the effects of speed on the Performance of Routing Protocols in Mobile Ad-hoc Networks. Corson et al.[10] discussed the Routing protocol in MANET with performance issues and evaluation considerations. Guangyu et.al. [11] presented the application layer routing as Fisheye State Routing in Mobile Ad Hoc Networks. In view of need to evaluate the performance of ODMRP with other common routing protocols used now days, simulation based experiments were performed by evaluating Packet Delivery Ratio, End to End delay and average throughput. Many researchers have evaluated multicast routing performance under a variety of mobility patterns [12-13].

The fisheye State Routing (FSR) algorithm for ad hoc networks introduces the notion of multi-level "scope" to reduce routing update overhead in large networks [14]. A node stores the link state for every destination in the network. It periodically broadcasts the link state update of a destination to its neighbors with a frequency that depends on the hop distance to that destination. Pei et al. [15] studied the routing accuracy of FSR and identified that it was comparable with an ideal Link State. FSR is more desirable for large mobile networks where mobility is high and the bandwidth is low. It has proved as a flexible solution to the challenge of maintaining accurate routes in ad hoc environments.

## 3. Experimental Setup

Evaluation of the performance of different routing techniques such as ODMRP, AODV and FSR was carried out through simulation using the GloMoSim v2.03 simulator [16]. The channel capacity of mobile hosts was set at 2Mbps. For each simulation, 60 nodes were randomly placed over a square field whose length and width is 1000 meters. Nodes communicate using MAC and CSMA for the routing protocols ODMRP, AODV and FSR. Each multicast source uses a Constant Bit Rate (CBR) flow. These parameters were chosen from "config.in" file within the simulator. Based on the requirements the values were adjusted and then it was executed. Monitored parameters were average throughput, end to end delay and packet delivery ratio (PDR).

## 4. Results and Discussion

The performance of the three routing protocols, i.e. ODMRP, AODV and FSR were evaluated under varying simulation conditions. The evaluation of performance was done on the basis of monitored parameters, average throughput, end to end delay and packet delivery ratio.

4.1 Average Throughput

Average throughput signifies the rate of packets communicated per unit time. The average throughput at a unit time (simulation time of 200 seconds) under varying number of nodes and mobility for all the simulated routing protocols are indicated in the Figure 1 (a-b). It can be observed that under most of nodal conditions the throughput of ODMRP is 4276.25 which are remarkably higher to throughput of AODV (3125.50) and throughput of FSR (487.25).

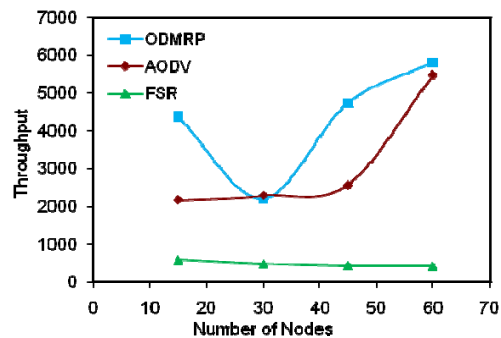

(a) under varying nodes

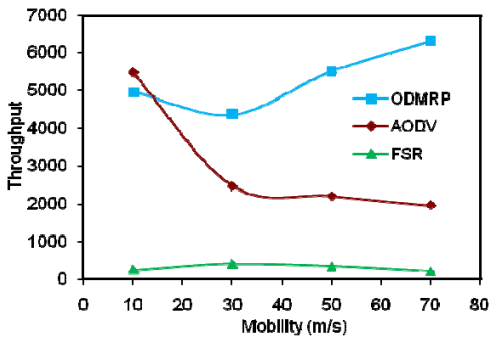

(b) under varying mobility

Figure 1 Average throughput under various input conditions

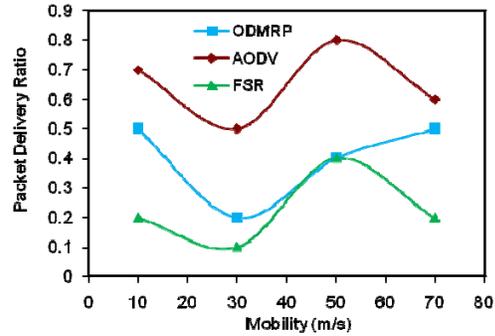

(b) under varying mobility

Figure 2 Packet delivery ratio under various input conditions

The FSR topology maintains up-to-date information received from neighboring nodes. The topology information is exchanged between neighbors via Unicast. Each node maintains network topology map for distance calculations and when network size increases, the amount of periodic routing information could become large. However the routing packets are not flooded. FSR captures pixels near the focal point with high detail. The details decrease as the distance from the focal point increase. When the mobility increases the routes to remote destinations become less accurate. The route table size still grows linearly with network size [14]. Hence throughput of FSR could here been lower than AODV and ODMRP.

Similarly for different mobility conditions too, ODMRP routing protocol displays increased performance as compared to the other two. The ODMRP average throughput with node mobility is 5276.75 bytes per simulation time as against AODV's 3024.00 and FSR's 298.75. The same reasons as stated for the improved performance of ODMRP under differing number of nodes can be given here too. The same behavior is experienced in the previous studies too under similar conditions [12].

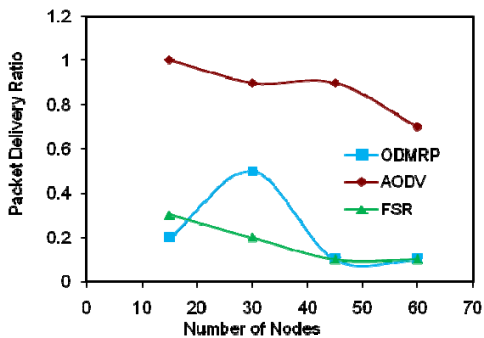

(a) under varying nodes

It can be observed that the PDR of AODV routing protocol is higher than the ODMRP and Fisheye state routing protocols. Higher the PDR, higher is the number of legitimate packets delivered without any errors. This shows that AODV exhibits a better delivery system as compared with the other two. The reasons for the higher PDR ratio of AODV can be attributed to its good performance in large networks with low traffic and low mobility. It discovers routes on-demand, and effectively uses available bandwidth. Also it is highly scalable and minimizes broadcast and transmission latency. Its efficient algorithm provides quick response to link breakage in active routes.

Moreover the ability of a routing algorithm to cope with the changes in routes is identified by varying the mobility. In this too the PDR of AODV protocol is higher as compared to the other two. The same reasons for the better PDR ratio of AODV under changing number of nodes can be given here too.

4.3 End-to-End Delay

The total latency between the source and destination experienced by a legitimate packet is given by end-to-end delay. It is calculated by summing up the time periods experienced as processing, packet, transmission, queuing and propagation delays. The speed of delivery is an important parameter in the present day competitive circumstances.

Higher end- to –end delay values imply that the routing protocol is not fully efficient and causes a congestion in the network. The values of end- to- end delay for the protocols ODMRP, AODV and FSR simulated at different number of nodes and differing mobility values are indicated in Figure 3. As against the other two protocols studied ODMRP exhibits lesser values of end-to-end delay. This implies that for ad hoc networks, the multicast routing protocol ODMRP exhibits a better performance than AODV and FSR.

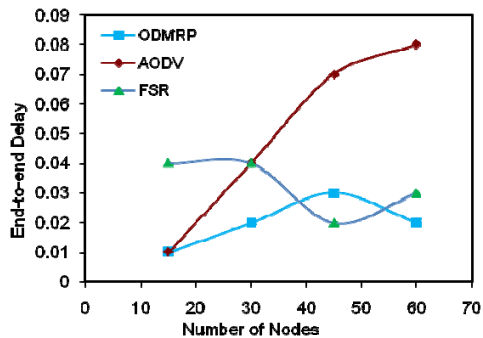

(a) under varying nodes

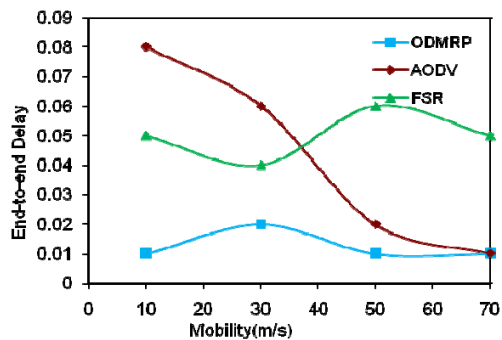

(b) under varying mobility

Figure 3 End-to-End Delay under various input conditions

## 5. Conclusions

Performance of the various routing protocols such as ODMRP, AODV and FSR were evaluated in this study. The following conclusions were drawn.

➢ Both under varying number of nodes and differing values of mobility Average throughput is higher for the routing protocol ODMRP. The maximum throughput of ODMRP is 43% higher than the maximum of AODV and FSR under varying nodes condition.

➢ AODV has a higher ratio of legitimate packet delivery as compared with the other routing protocols evaluated, ODMRP and FSR. The maximum packet delivery of AODV is 38% higher than the maximum of ODMRP and FSR under varying nodes condition.

➢ ODRMP performs better in avoiding network congestion as compared to AODV and FSR. The better your paper looks, the better the Journal looks. Thanks for your cooperation and contribution.

## Acknowledgments

The authors gratefully acknowledge Dr.P.Chinnadurai, Secretary & Correspondent, Panimalar Engineering College, Chennai for his encouragement and continuous support. They also appreciate the help rendered by Dr.L.Karthikeyan.